\documentclass[pra, twocolumn,amsfonts,amsmath,amssymb,eufrak]{revtex4-2}

\usepackage{epsfig}
\usepackage{color}
\usepackage{bm}
\usepackage{hyperref}
\usepackage[hyphenbreaks]{breakurl}

\usepackage{braket}
\usepackage{graphicx}
\usepackage{amsthm}

\newcommand{\eq}[1]{\begin{equation} #1 \end{equation}}
\newcommand{\eqa}[2]{\begin{equation} #1 \label{#2} \end{equation}}
\newcommand{\balign}[1]{\begin{align} #1 \end{align}}

\newcommand{\mx}[1]{\begin{pmatrix}#1 \end{pmatrix}}

\newcommand{\figin}[4]
{\begin{figure}[tb]
\centering
\includegraphics[width= #1]{#2.pdf}
\caption{#3}
\label{f:#4}
\end{figure}}

\newcommand{\todayd}{\the\year/\the\month/\the\day}

\newcommand{\bib}{\bibitem}

\newcommand{\lmd}{\lambda}

\newcommand{\nt}{\notag}

\newcommand{\bel}{\begin{easylist}}
\newcommand{\eel}{\end{easylist}}

\newcommand{\eref}[1]{Eq.~\eqref{#1}}

\def \({\left(}
\def \){\right)}
\def \[{\left[}
\def \]{\right]}

\newcommand{\abs}[1]{\left|#1\right|}

\newcommand{\sumtwo}[2]%
{\mathop{\sum_{#1}}_{#2}}
\newcommand{\sumthree}[3]%
{\mathop{\mathop{\sum_{#1}}_{#2}}_{#3}}
\newcommand{\sumfour}[4]%
{\mathop{\mathop{\mathop{\sum_{#1}}_{#2}}_{#3}}_{#4}} 
\newcommand{\prodtwo}[2]%
{\mathop{\prod_{#1}}_{#2}}
\newcommand{\mintwo}[2]%
{\mathop{\min_{#1}}_{#2}}
\newcommand{\maxtwo}[2]%
{\mathop{\max_{#1}}_{#2}}
\newcommand{\maxthree}[3]%
{\mathop{\mathop{\max_{#1}}_{#2}}_{#3}}
\newcommand{\limtwo}[2]%
{\mathop{\lim_{#1}}_{#2}}
\newcommand{\suptwo}[2]%
{\mathop{\sup_{#1}}_{#2}}
\newcommand{\supthree}[3]%
{\mathop{\mathop{\sup_{#1}}_{#2}}_{#3}}
\newcommand{\supfour}[4]%
{\mathop{\mathop{\mathop{\sup_{#1}}_{#2}}_{#3}}_{#4}} 
\newcommand{\inftwo}[2]%
{\mathop{\inf_{#1}}_{#2}}
\newcommand{\infthree}[3]%
{\mathop{\mathop{\inf_{#1}}_{#2}}_{#3}}
\newcommand{\inffour}[4]%
{\mathop{\mathop{\mathop{\inf_{#1}}_{#2}}_{#3}}_{#4}} 

\newcommand\calN{{\cal N}}

\newcommand{\bsp}{\boldsymbol{p}}

\newcommand{\ep}{\varepsilon}

\newcommand{\Di}{\mathit{\Delta}}

\newcommand{\para}[1]{{\em #1}\/.---}

\newtheorem{thm}{Theorem}
\newtheorem{lm}[thm]{Lemma}
\newtheorem{pro}[thm]{Proposition}

\theoremstyle{definition}
\newtheorem{dfn}[thm]{Definition}

\def\rnum#1{\resizebox{0.5em}{\height}{\expandafter{\romannumeral #1}}}
\def\Rnum#1{\resizebox{0.5em}{\height}{\uppercase\expandafter{\romannumeral #1}}}

\newcommand{\sgmss}{\sigma^{\rm ss}}
\newcommand{\pss}{p^{\rm ss}}

\def \A{{\rm A}}
\def \B{{\rm B}}
\def \C{{\rm C}}
\newcommand{\hp}{\hat{p}}
\newcommand{\hR}{\hat{R}}
\renewcommand{\Re}{{\rm Re}}
\newcommand{\lmdR}{\lmd_{\rm R}}
\newcommand{\lmdI}{\lmd_{\rm I}}

\begin{document}


\title{
Second largest eigenvalue does not bound stationary entropy production
}

\author{Naoto Shiraishi} 
\email{shiraishi@phys.c.u-tokyo.ac.jp}
\affiliation{Department of Basic Science, The University of Tokyo, 3-8-1 Komaba, Meguro-ku, Tokyo 153-8902, Japan}%



\begin{abstract}
We examine the spectral dissipation-coherence trade-off inequality conjectured by Oberreiter {\it et al}. [Phys. Rev. E 106, 014106 (2022)], stating that the coherent number, defined from the second largest eigenvalue, provides a lower bound on the stationary entropy production per unit oscillation.
We disprove this conjecture by explicitly constructing a counterexample, which accompanies finite coherent number with vanishing stationary entropy production.
This model also excludes a wide class of thermodynamic bound with the real and imaginary parts of the second largest eigenvalue of the transition rate matrix.
Our result implies that the second largest eigenvalue does not necessarily characterize the degree of coherent oscillation.
\end{abstract}

\maketitle

\para{Introduction}
Entropy production is a key quantity in nonequilibrium stochastic systems~\cite{Shibook, Seibook}.
A standard characterization of entropy production is the degree of irreversibility of thermodynamic processes.
The entropy production plays a crucial role in the nonlinear response theory including the MacLennan-Zubarev expression~\cite{ZK70} and transient time correlation function method developed by Evans and Morriss~\cite{EMbook}.
In 1990's, the celebrated fluctuation theorem uncovered a hidden symmetry in entropy production, thereby characterizing thermodynamic irreversibility through a quantitative relation~\cite{ECM93, Jar97, Kur98, Jar00}.
Subsequent studies determine the origin of entropy production by decomposing it at the level of subsystems and single transitions~\cite{SU12, SS15}.

In the last decade, the entropy production has been recognized also as a lower bound in a variety of trade-offs in nonequilibrium dynamics.
A representative example is the thermodynamic speed limit inequality, which prohibits a quick thermodynamic process with small entropy production~\cite{SFS18, VVH20, Dec22, VS23, Shi24}.
A related relation provides a universal trade-off relation between power and efficiency of heat engines~\cite{SST16, SS19}.
Another important achievement is the thermodynamic uncertainty relation, where the stationary entropy production is a thermodynamic cost to suppress a relative current fluctuation~\cite{BS15, Ging16, GRH17, DS20, Shi21}.
Moreover, a stronger bound than the second law in relaxation processes has been derived, which limits possible relaxation paths~\cite{SS19b, Kol22}.

Recently, motivated by oscillatory phenomena in biological systems, the connection between stationary entropy production and the degree of coherent oscillation has been actively studied~\cite{Cao15, BS17}.
The thermodynamic cost of autocorrelation functions (oscillation) itself has long been studied~\cite{TT74, TS08}, while recent work has highlighted the importance of the competition of oscillation and decaying speed, leading to a deeper understanding from this perspective.
In particular, the spectral dissipation-coherence trade-off (DCT) inequality proposed by Oberreiter {\it et al}.~\cite{OSB22} has stimulated extensive research and played a central role in subsequent developments in this field.
The spectral DCT inequality claims that the stationary entropy production per unit oscillation is bounded from below by the coherent number, where the coherent number and oscillation are quantified by using the real and imaginary parts of the second largest eigenvalue of the transition rate matrix of the Markov jump process.
Oberreiter {\it et al}. conjectured the universality of this spectral DCT inequality on the basis of various numerical simulations and analyses on several tractable toy models.
After this proposal, various studies have tried to prove this conjecture and several interesting inequalities relating thermodynamic dissipation and coherent oscillation have been found~\cite{OIK23, Shi23, KOI24, VVS24, Gu24}, though the original inequality has still been left unresolved.

The validity of the spectral DCT inequality remains unsettled.
On the one hand, the spectral DCT inequality has been proven in some specific settings.
Santolin and Falasco~\cite{SF25} proposed another DCT inequality with the correlation time and mean oscillation velocity mainly for overdamped Langevin systems with limit cycles.
This DCT inequality, including its extension to Markov jump processes, was proved in the weak noise limit and for asymptotically Gaussian oscillations~\cite{NI25, Kol26}.
In the former limit, this DCT inequality coincides with the spectral DCT inequality.
Another progress is seen in the linear Langevin systems, where Nagayama {\it et al}.~\cite{NKI26} proved a stronger inequality than the spectral DCT.
On the other hand, Gu~\cite{Gu26} derived a weaker version of the spectral DCT inequality involving an eigenvector-dependent factor and argued that straightforward applications of standard inequality techniques do not recover the original spectral bound. 
On the basis of these observations, it is argued that the original bound may be violated when the dominant oscillatory mode is strongly localized, although no explicit counterexample was provided.
Very recently, a candidate of a counterexample exhibiting a numerical violation of the spectral DCT inequality was proposed, while the author is cautious about the conclusion~\cite{blog26}.

In this Letter, we put a decisive answer to this controversy.
We disproves the conjectured DCT inequality by explicitly constructing a counterexample to the spectral DCT inequality, where the violation is analytically provable.
Our model has finite real and imaginary parts of the second largest eigenvalue with vanishing stationary entropy production.
Beyond disproving the conjecture, this model also rules out the existence of a broader class of inequalities bounding the stationary entropy production by the second largest eigenvalue.
Our result clearly shows that the second largest eigenvalue does not serve as a universal lower bound on the thermodynamic dissipation.

\para{Setting and conjectured relation}
We consider a Markov jump process on discrete states.
The time evolution of the probability distribution $\bsp$ is given by the master equation
\eq{
\frac{d}{dt}\bsp=R\bsp,
}
where $R$ is the transition rate matrix, satisfying nonnegativity $R_{ij}\geq 0$ for $i\neq j$ and the normalization condition $\sum_i R_{ij}=0$.
We suppose that this $R$ accompanies a unique stationary distribution $\bsp^{\rm ss}$.

We now formulate stochastic thermodynamics in this setting.
Assuming the local detailed-balance condition, its stationary entropy production is given by
\eq{
\sgmss:=\sum_{i,j}R_{ij}\pss_j \ln \frac{R_{ij}\pss_j}{R_{ji}\pss_i},
}
which quantifies the thermodynamic cost of this system per unit time~\cite{Shibook}.

Now we investigate the stationary oscillation of this system, which can be observed as oscillatory relaxation following a fluctuation away from the stationary value.
To quantify the strength of oscillation, however, one needs to take into account not only how fast the system oscillates but also how slowly the fluctuation decays.
For this reason, the ratio of the absolute values of the real part $\lmdR$ and the imaginary part $\lmdI$ of the second largest eigenvalue is considered as a quantifier of oscillations. 
Here, the second largest eigenvalue $\lmd_2$ denotes the eigenvalue of the transition rate matrix $R$ with the second-largest real part. 
The decay rate of fluctuations is given by its minus of the real part, whereas the angular frequency of the oscillation is given by its imaginary part.

To capture the thermodynamic cost of oscillation, Oberreiter, Seifert, and Barato~\cite{OSB22} conjectured the following spectral dissipation-coherent trade-off (DCT) inequality
\eqa{
\Di S\geq 4\pi^2 \calN
}{OSB}
for $\calN\geq (2\pi)^{-1}$, where $\Di S:=2\pi\sgmss/\lmdI$ is the stationary entropy production per single oscillation and $\calN:=\lmdI/2\pi \lmdR$ is coherent number quantifying the quality of oscillation.
The spectral DCT inequality is thus same as the following lower bound for stationary entropy production
\eq{
\sgmss\geq \frac{\lmdI^2}{\lmdR}.
}
The condition $\calN\geq (2\pi)^{-1}$ reads $\lmdI\geq \lmdR$, meaning that the decaying is slow compared to oscillation.
The spectral DCT inequality \eqref{OSB} is supported by various numerical simulations~\cite{OSB22} and is proven in the weak-noise regime~\cite{SF25, NI25, Kol26} and in the linear Langevin systems~\cite{NKI26}.

\para{Counterexample to dissipation-coherent trade-off inequality}
Contrarily to these expectations, we provide a counterexample to the spectral DCT inequality \eqref{OSB}.
Precisely, we construct a model accompanying finite $\lmdR$ and $\lmdI$, while its stationary entropy production goes to zero; $\sgmss\to 0$.
Hence, this model disproves not only the conjectured spectral DCT inequality \eqref{OSB} and its constant-factor variants $\Di S\geq c \calN$ with a constant $c$, but also any inequality of the following general form:
\eqa{
\sgmss \geq C\frac{\lmdI^\alpha}{\lmdR^\beta} \hspace{10pt} \text{[false inequality]}
}{gen}
with positive constants $\alpha,\beta,C>0$.
This clearly shows that the second eigenvalue does not serve as a universal lower bound of the stationary entropy production in a general setting.

\figin{8.7cm}{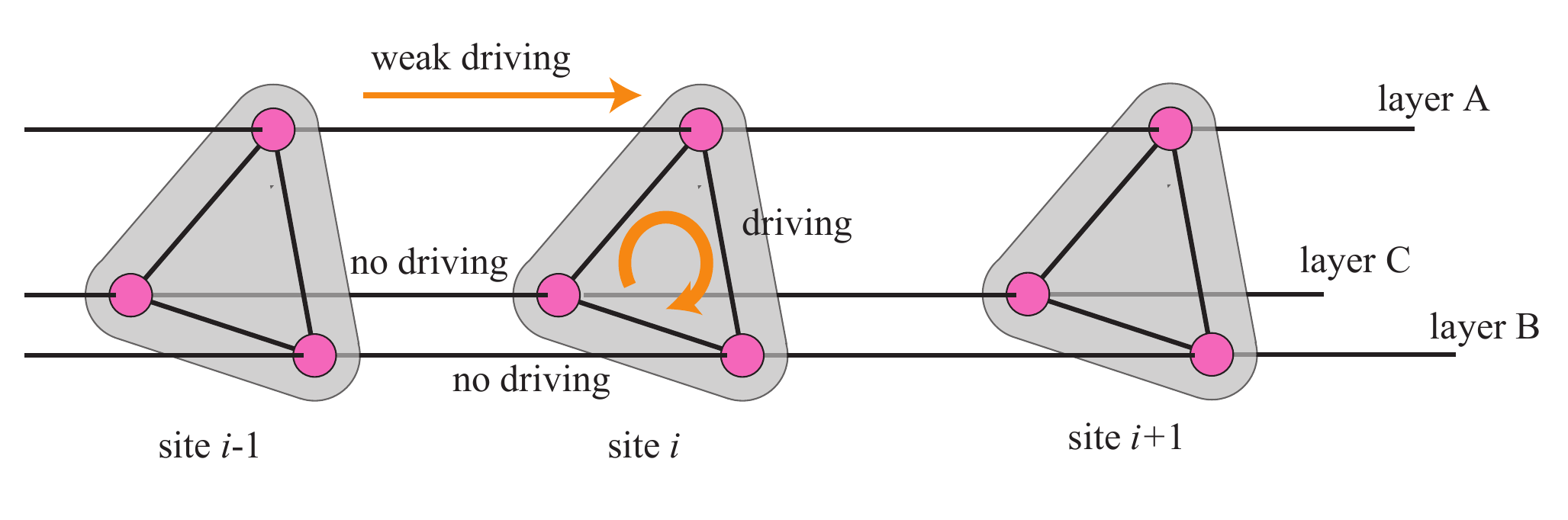}{
Schematic of our model disproving the spectral DCT inequality.
Our model consists of three layers, A, B, and C, which are one-dimensional with $L$ sites with the periodic boundary condition.
The state of the system is specified by $(i,a)$ with site $1\leq i\leq L$ and layer $a\in \{ \A, \B, \C\}$.
In a single jump, the site changes as $i\to i\pm 1$ in  the same layer, or the layer changes in the same site.
Layer A has small nonequilibrium driving, while layers B and C have no driving.
In a single site, we have nonequilibrium driving in the direction $\A\to \B\to \C\to \A$.
The stationary probability of layer A is extremely small (of order $O(\ep^2)$).
}{schematic}

\para{Model construction}
Our model consists of $3L$ states arranged in three one-dimensional layers, A, B, and C, each of which contains $L$ states.
A state is labeled as $(i,a)$ with site label $1\leq i\leq L$ and layer label $a\in \{ \A, \B, \C\}$.
Our model has two types of transitions: transitions within the same layer that change $i\to i\pm1$, and transitions between layers at the same site $i$, both of which are shift invariant.

The transition rate in layer A is weakly driven as
\eq{
r^\pm_\A=\frac{\ep}{2(1-\cos \theta)}\pm \frac{1}{\sin \theta}
}
and those in layer B and C have no driving
\eq{
r^\pm_\B=r^\pm_\C=\frac{K}{2(1-\cos \theta)}
}
with $\theta=2\pi/L$, where $\ep=1/L^q$ is a small parameter with $1/2<q<1$ and $K$ is a large but finite constant.
For our purpose, $K=10$ suffices to prove our desired properties.
The condition $q<1$ confirms the nonnegativity of transition rate $r_->0$.
We employed the periodic boundary condition and identified site $L+1$ to site 1.

The transition rate matrix $Q$ among different layers is given by
\eq{
Q=Q_0+Q_1
}
with
\balign{
Q_0&= \mx{-a &u &u \\
a/2 &-u-h& h \\
a/2&h&-u-h
}
,\hspace{10pt}
Q_1=\mx{
0&-v&v \\
b&0&-v \\
-b&v&0
},
}
where we set $a=1-\ep$, $u=a\frac{p}{1-p}$ with $p=O(\ep^2)$, and $h=(a+d\ep)/2$ with a constant $d>2$.
We also set $v$ and $b$ as $v=(d-1)\ep\sqrt{p}$($=O(\ep^2)$) and $b=(d-1)(1-p)\ep/2\sqrt{p}$($=O(1)$) so that we have $b=(1-p)v/2p$ and $v(2b+v)=(d-1)^2\ep^2$.
In addition, in order to satisfy the nonnegativity of the off-diagonal elements, which is equivalent to $a\sqrt{p}\geq (d-1)\ep (1-p)$, we require $p>(d-1)\ep^2$.
The matrix $Q_0$ provides its equilibrium shape, and $Q_1$ provides weak nonequilibrium driving.

Owing to this elaborated construction, we can compute both its stationary distribution and its eigenvalues explicitly.
First, it is easy to confirm that both $Q_0$ and $Q_1$ have the same stationary distribution
\eq{
\bsp^{\rm ss}_Q=\mx{p\\ \frac{1-p}{2} \\ \frac{1-p}{2}},
}
and thus $Q$ also has $\bsp^{\rm ss}_Q$ as its stationary distribution.
In addition, the characteristic polynomial of $Q$ is computed as
\balign{
&\det (zI-Q) \nt \\
=&z\[ \( z+\frac{1-\ep}{1-p}+\frac d2 \ep\)^2+ \(  (d-1)^2- \frac{d^2}{4}\) \ep^2\] ,
}
which implies the nonzero eigenvalues of $Q$ as
\eqa{
\lmd_\pm=-\frac{1-\ep}{1-p}-\frac d2 \ep \pm i\sqrt{(d-1)^2- \frac{d^2}{4}}\ep.
}{lmd-R0}
Note that the condition $d>2$ confirms that the term in the square root is positive, $(d-1)^2- \frac{d^2}{4}>0$, and its real part $-1-(d/2-1)\ep+O(\ep^2)$ is shifted below $-1$ already at first order in $\ep$.

\para{Evaluating the second largest eigenvalue}
We first compute the second largest eigenvalue of the transition rate matrix $R$ of the entire system.
To this end, we employ the inverse Fourier transform with respect to the site index:
\eq{
p_{x,a}(t)=\frac1L \sum_{k=0}^{L-1}e^{2\pi i kx/L}\hp_{k,a}(t),
}
with which the transition rate matrix $R$ is block-diagonalized as diag$(\hR^0, \hR^1,\ldots , \hR^{L-1})$.
Here, $\hR^k$ is a $3\times 3$ matrix given by 
\eq{
\hR^k=Q-F_k\mx{\ep&& \\ &K& \\ &&K}+2iG_k\mx{1&& \\ &0& \\ &&0}
}
with $F_k:=({1-\cos \frac{2\pi k}{L}})/ ({1-\cos \frac{2\pi}{L}})$ and $G_k:={\sin \frac{2\pi k}{L}}/{\sin \frac{2\pi}{L}}$.
This matrix describes the time evolution of the mode-$k$ probability distribution:
\eq{
\frac{d}{dt}\mx{\hp_{k,\A} \\ \hp_{k,\B} \\ \hp_{k,\C}}=\hR^k \mx{\hp_{k,\A} \\ \hp_{k,\B} \\ \hp_{k,\C}}.
}
We denote three eigenvalues of $\hR^k$ by $\lmd^k_1, \lmd^k_2, \lmd^k_3$ ($\Re \lmd^k_1\geq \Re \lmd^k_2\geq \Re \lmd^k_3$).
We note that $\hR^k$ and $\hR^{L-k}$ are complex conjugates of each other, and thus their eigenvalues are also complex conjugates of each other.

We now determine the eigenvalues of $\hR^k$ for each $k$.
In the case of $k=0$, $\hR^0$ is equal to $Q$, and thus its three eigenvalues are $\lmd^0_1=0$ and $\lmd^0_2, \lmd^0_3=\lmd_\pm$ given in \eref{lmd-R0}.

To evaluate the eiganvalues of $\hR^1$, we apply the Gershgorin circle theorem~\cite{Bhabook}.
The first Gershgorin disc is centered at $-1+2i$ and has radius $\abs{u-v}+\abs{u+v}=O(\ep^2)$.
The other two discs are centered approximately at $-K$ and have radii less than 2, having no overlap to the first disc.
Consequently, the largest eigenvalue of $\hR^1$ settles in the first disc and $\Re \lmd^1_1=-1+O(\ep^2)$.
The same argument holds for $\hR^{L-1}$.

We shall show that all the other eigenvalues of $\hR^k$ for $2\leq k\leq L-2$ have their real parts less than $-1+O(\ep^2)$.
Noticing $F_k\geq F_2=4\cos^2 \frac{\pi}{L} > 3$ for these $k$, we find that the real parts of the centers of their first Gershgorin discs are less than $-1-2\ep+\ep^2$ and their radii are $O(\ep^2)$. 
The other two discs are centered approximately $-F_kK<-3K$ and have radii less than 2.

In summary,comparing $\Re \lmd^0_2=-1-(d/2-1)\ep+O(\ep^2)$, $\Re \lmd^1_1=\Re \lmd^{L-1}_1=-1+O(\ep^2)$, and $\Re \lmd^k_1\leq -1-2\ep+O(\ep^2)$ for $2\leq k\leq L-2$, we conclude that the second largest eigenvalue of $R$ is $\lmd^1_1$ (and $\lmd^{L-1}_1$), which satisfies $\lmdR=\abs{-1+O(\ep^2)}=O(1)$, $\lmdI=2+O(\ep^2)=O(1)$, and $\lmdI\geq \lmdR$.

\para{Evaluating the stationary entropy production}
We next evaluate the amount of the stationary entropy production rate $\sgmss$.
Since the layers B and C have no driving, the dissipation comes from (i) nonequilibrium flow in layer A, and (ii) internal dissipation among layers A, B, and C in a single site.
We compute the orders of these two contributions in $\ep$ and $L$.

The contribution of (i) is given by
\eq{
\sigma_\A=p(r_+-r_-)\ln \frac{r_+}{r_-}.
}
Since $\theta=O(1/L)$ and $p=O(\ep^2)$, we find that $r_\pm=O(\ep L^2)\pm O(L)$ and thus have
\eq{
\sigma_\A=O(\ep^2 L)\ln (1+O((\ep L)^{-1}))=O(\ep),
}
where we used $(\ep L)^{-1}=O(L^{-1+q})$ is a small quantity.
Taking $L\to \infty$ limit, $\sigma_\A$ vanishes.

We next evaluate the contribution of (ii) in each site.
We notice that the stationary probability current $\A\to \B\to \C\to \A$ in each site is $2pb/L=O(\ep^2/L)$ and the strength of nonequilibrium driving among layers is $O(1)$.
Thus, the contribution (ii) over all $L$ sites is computed as
\eq{
\sigma_{\rm in}=LO\( \frac{\ep^2}{L}\)=O(\ep^2)
}
which again goes to zero in the $L\to \infty$ limit.

In summary, stationary entropy production rate $\sgmss$ is $O(\ep)$, which vanishes in the $\ep\to 0$ limit.
Combining the evaluation of the second largest eigenvalue, we conclude that this model is indeed a counterexample to the spectral DCT inequality \eqref{OSB} and any thermodynamic bound of the form \eqref{gen}.

\para{Ideas behind this construction}
The key idea of this construction is to decompose the second largest eigenvalue and the dominant dissipative dynamics, by employing a rare oscillating layer and equilibrium layers as \cite{blog26}.
The second largest eigenvalue of $R$, $\lmd^1_1$, describes the dynamics of mode $k=1$ mainly in layer A.
On the other hand, since the stationary weight on layer A is $p=O(\ep^2)$, which accompanies negligible contribution to the stationary entropy production.
This dynamics is characterized by $\lmd^0_2$, which is not the second largest but the third largest eigenvalue of $R$.

To realize this situation in a controlled manner, we carefully tune the parameters.
The nonequilibrium driving in layer A, $r_+-r_-$, is large ($O(L)$), while the transition rate in each layer itself is larger ($O(\ep L^2)$).
This choice makes the eigenvalue of the $k=1$ mode well separated from all the other eigenvalues, with keeping the thermodynamic affinity $\ln (r_+/r_-)$ small ($O((\ep L)^{-1})$).

The nonequilibrium driving among layers is also necessary for the above construction.
In fact, if we set $Q_1=O$, then $Q$ has eigenvalues 0, $-a-u=-\frac{1-\ep}{1-p}$, $-u-2h$, and the second largest eigenvalue $\lmd^0_2=-\frac{1-\ep}{1-p}=-1+\ep-O(\ep^2)$ exceeds $\Re \lmd^1_1$.
In general, if no nonequilibrium driving among layers is applied, the Gershgorin disc for $\lmd^1_1$ extends to the left of $\lmd^0_2$.
To shift $\Re \lmd^0_2$ below $-1$, we add the perturbation $Q_1$, which mixes the corresponding mode with another mode $-u-2h$, which is also close to $-1$.
This is why our counterexample consists of three layers, not two layers.

\para{Implications of this result}
Our counterexample disproves the conjectured spectral DCT inequality.
This fact, however, does not suggest that the stationary oscillation does not bound the entropy production.
Rather, our counterexample elucidates the discrepancy between the coherent oscillation and the second largest eigenvalue.
These two are usually identified since the slowest relaxation of the time correlation function is described as
\eq{
A e^{-\lmdR t} \sin (\lmdI t)+\cdots .
}
However, this description does not guarantee that its amplitude $A$ is nonnegligible, and our counterexample indeed has vanishing amplitude $A$ on this mode.
Namely, our model has a very rare but strong coherent mode, and thus the second largest eigenvalue itself cannot bound the stationary entropy production.
This observation is in line with the argument by Gu~\cite{Gu26}, suggesting that the OSB inequality might be violated when the second eigenmode has only a small overlap with the stationary distribution.

Nevertheless, there is also good reason to focus on the second largest eigenvalue: in many oscillator models, it indeed corresponds to a dominant oscillatory mode.
In fact, the above correspondence is true for a variety of systems and settings~\cite{OSB22, SF25, NI25, Kol26, NKI26}.
In such cases, any attempt to suppress the weight of the second eigenmode in the stationary distribution comes at an additional cost, preventing a violation of the spectral DCT inequality.
However, this compensation mechanism is not universal, and our counterexample indeed achieves a highly biased stationary distribution without paying any cost.

It is worth comparing our counterexample and a provable DCT inequality in the homogeneous linear Langevin systems shown by Nagayama {\it et al}.~\cite{NKI26}.
First, in the setting of Nagayama {\it et al}. the homogeneity of the system prohibits the coexistence of a rare driven layer and an equilibrium layer.
In addition, the asymmetry of generator and the amount of stationary entropy production is tightly connected in the linear Langevin systems, which is a key ingredient in the proof of Nagayama {\it et al}..
On the other hand, this connection no longer holds in general Markov jump processes, and our counterexample possesses highly asymmetric generator and vanishing stationary entropy production.

\para{Discussion}
We settle the DCT conjecture raised in Ref.~\cite{OSB22} in the negative by explicitly constructing a counterexample.
This counterexample excludes a wide class of bound with the second largest eigenvalue given in the form of \eref{gen}.
More precisely, for any inequality \eqref{gen} with given $\alpha$, $\beta$, and $C$, we have a counterexample to this inequality with a finite $L$.

Our results point to two promising directions for future work. 
First, it may be fruitful to characterize coherent oscillations using quantities other than the second eigenvalue and to seek thermodynamic bounds in terms of such quantities. 
Previous studies have revealed that eigenvectors~\cite{Gu26} and oscillating observables themselves~\cite{Shi23} serve as universal lower bounds of stationary entropy production.
Refinement of these bounds to a more tractable and tight one will be a hopeful approach.

A complementary direction is to restrict the class of Markov jump processes under consideration and establish the original spectral DCT inequality within such a restricted setting. 
The DCT inequality \eqref{OSB} can be proven in the weak-noise regime~\cite{OSB22, SF25, NI25, Kol26}, and a stronger inequality than \eref{OSB} can also be proven in the homogeneous linear Langevin systems~\cite{NKI26}.
Observing them, in many cases the second largest eigenvalue appears to describe the dominant oscillatory mode.
Clarifying when the second largest eigenvalue is indeed dominant merits a future investigation.

\para{Note added}
When I finished this manuscript, I noticed a 12-state numerical counterexample to the spectral DCT inequality~\cite{Jas26}.
As far as I see, the construction of this counterexample is based on different ideas from ours.

{\it Acknowledgments.} ---
The author used ChatGPT-6 Astra and ChatGPT-5.6 Sol for finding a counterexample and improving and polish the English expressions in the manuscript.
This work was supported by Grant-in-Aid for Early-Career Scientists 26K17048, Grant-in-Aid for Transformative Research Areas (B) 26K00021, and JST ERATO Grant Number JPMJER2302.



\end{document}